\documentclass[aps,prl,twocolumn,superscriptaddress,showpacs]{revtex4}
\usepackage{graphics,epsfig,amsmath,amssymb}
\def\ket#1{\vert#1\rangle}
\def\ketbra#1{\vert#1\rangle\langle#1\vert}
\begin{document}

\title{Remote state preparation: arbitrary remote control of photon polarization}

\author{Nicholas A. Peters}
\affiliation{Physics Department, University of Illinois, 1110 West Green Street, Urbana, IL 61801} 
\author{Julio T. Barreiro}
\affiliation{Physics Department, University of Illinois, 1110 West Green Street, Urbana, IL 61801}
\author{Michael E. Goggin}
\affiliation{Physics Department, University of Illinois, 1110 West Green Street, Urbana, IL 61801}
\affiliation{Physics Department, Truman State University, Kirksville, MO 63501}
\author{Tzu-Chieh Wei}
\affiliation{Physics Department, University of Illinois, 1110 West Green Street, Urbana, IL 61801}
\author{Paul G. Kwiat}
\affiliation{Physics Department, University of Illinois, 1110 West Green Street, Urbana, IL 61801}

\date{\today}

\begin{abstract}
We experimentally demonstrate the first remote state preparation of arbitrary single-qubit states, encoded in the polarization of photons generated by spontaneous parametric downconversion.  Utilizing degenerate and nondegenerate wavelength entangled sources, we remotely prepare arbitrary states at two wavelengths.  Further, we derive theoretical bounds on the states that may be remotely prepared for given two-qubit resources. 
\end{abstract}

\maketitle
Quantum communication is concerned with the transmission, manipulation, and detection of quantum information.  If a sender (Alice) wants to transmit  an {\it unknown} quantum state to a receiver (Bob), they may use teleportation~\cite{tele}.  However, it has been shown that the classical communication costs for sending a {\it known} state using the remote state preparation protocol (RSP) are less than those of teleportation~\cite{lo00,pati00,bennett01}.  RSP is a quantum communication protocol that relies on correlations between two entangled qubits to similarly prepare Bob's qubit in a particular state determined by Alice, conditional on the outcome of a measurement on her qubit.  However, unlike teleportation, RSP does not require the sender to perform full Bell-state analysis, currently an experimental challenge for optical implementations.

Thus far, several RSP demonstrations with varying degrees of control over remotely prepared qubits have been reported:  pseudo-pure states using liquid-state NMR~\cite{peng03a}, pure-state superpositions of vacuum and single-photon states~\cite{bbl04}, and some mixed states of a polarization qubit~\cite{jpk04,ericsson}.  However, until now, no RSP implementation has achieved control over the three parameters required to prepare {\em arbitrary} single qubit states, which we report here.  Specifically, we achieve arbitrary mixed state RSP by using arbitrary polarization measurement on one photon of a polarization-entangled pair.  In addition, we derive bounds on the states that may be remotely prepared using arbitrary two-qubit entangled resources and discuss two specific cases in detail.  

First, we describe the general idea of RSP and give several examples.  Although we will make reference to photon polarization qubits, the methods described here can be generalized to any physical qubit implementation.  Consider the two-photon maximally entangled state:  
$|\phi^+\rangle \equiv (|H_{t}H_{rp}\rangle+|V_{t}V_{rp}\rangle)/
\sqrt{2} \equiv (|D_{t}D_{rp}\rangle+|A_{t}A_{rp}\rangle)/
\sqrt{2}$, where the subscripts label the trigger and remotely 
prepared photons, $|H\rangle$ and 
$|V\rangle$ label horizontal and vertical polarization states and 
$|D\rangle\equiv(|H\rangle+|V\rangle)/\sqrt{2}$ and 
$|A\rangle\equiv(|H\rangle-|V\rangle)/\sqrt{2}$ label diagonal 
and anti-diagonal polarizations states, respectively.  Measurement of the trigger photon in the state $|D_t\rangle$  (i.e., detecting the trigger 
photon after a diagonal polarizer) prepares the other photon in the state $|D
_{rp}\rangle$.  To remotely prepare an arbitrary {\it pure} state $|\psi_{rp}(\theta,\phi)\rangle\equiv\cos\theta|D\rangle+\sin\theta e^{i\phi}|A\rangle$, Alice can act on the trigger photon with a quarter-wave plate (QWP) and a half-wave plate (HWP), such that the two-photon state $|\phi^+\rangle \rightarrow(|\zeta_{t}(\theta,\phi) D_{rp}\rangle+|\zeta_{t}^{\bot}(\theta,\phi) A_{rp}\rangle)/\sqrt{2}\equiv(|D_{t} \psi_{rp}(\theta,\phi)\rangle+|A_{t} \psi_{rp}^{\bot}(\theta,\phi)\rangle)/\sqrt{2}$, where $|\zeta_{t}(\theta,\phi)\rangle\equiv\cos\theta|D\rangle-e^{-i\phi}\sin\theta|A\rangle$, and $\langle\zeta^{\bot}|\zeta\rangle=0$.  
Thus when the trigger qubit is projected into $\langle D|$ ($\langle A|$), the remotely prepared qubit is in the state $|\psi(\theta,\phi)\rangle$ ($|\psi^{\bot}(\theta,\phi)\rangle$).  The 50\% efficiency in this case can be improved to 100\% if the state Alice is sending is constrained to lie on a single great circle on the Poincar\'e sphere:  Bob simply performs the appropriate transformation on his photon $|\psi_b^{\perp}\rangle\rightarrow |\psi_b\rangle$ whenever Alice reports that she detects her photon in the state $|A\rangle$.  This procedure does {\it not} work in general due to the impossibility of a universal NOT operation on arbitrary qubit states~\cite{BHW99}.  

If instead, the trigger polarizer 
is removed, the trigger photon is measured in a polarization-{\it insensitive} 
way, tracing over its polarization state. This prepares the remaining photon in the completely mixed state 
(i.e., unpolarized), according to
\begin{equation}
\begin{array}{c}
\left. \rho_{rp} = \langle D_t|\phi^+\rangle \langle\phi^+|D_t
\rangle+\langle A_t|\phi^+\rangle \langle\phi^+|A_t\rangle =  \right
. \\
\frac{1}{2}(|D_{rp}\rangle \langle D_{rp}|+|A_{rp}\rangle \langle A_{
rp}|) =  \frac{1}{2}\left( \begin{array}{cc}1 & 0 \\
0 & 1 \end{array} \right).
\label{rhomixed} 
\end{array}
\end{equation}
By using a {\it partial} polarizer to tune between the two limiting cases discussed above, we can control the {\it strength} of the 
polarization measurement on the trigger, and thus the resulting mixedness of 
the remotely prepared qubit (RPQ).  Combined with the wave plates that allow us to prepare arbitrary pure states, the partial polarizer allows us to prepare completely arbitrary mixed states:
\begin{equation}
\rho_{rp}((\theta,\phi,\lambda))=(1-\lambda)|\psi(\theta,\phi)\rangle\langle\psi(\theta,\phi)|+\frac{\lambda}{2}\openone,
\label{arbitqubit}
\end{equation}
where the value $\lambda$ is determined by the partial polarizer.  

The experiment divides into three logical sections:  entangled resource state creation, detection of the trigger to remotely prepare a qubit, and tomography of the RPQ.  Photons are created via spontaneous parametric downconversion (SPDC) by pumping two type-I phasematched $\beta$-Barium Borate (BBO) nonlinear crystals with a cw Ar-ion 351-nm pump laser.  The two crystals have their optic axes oriented in perpendicular planes to give an entanglement superposition $(|HH\rangle+|VV\rangle)/\sqrt{2}$~\cite{Kwiat99}.  An initial tomography of the entangled resource state is taken by measuring 36 polarization correlations (such as $|HH\rangle$, $|HV\rangle$, $|DH\rangle$,..., etc.) from which a density matrix is calculated using a maximum likelihood technique~\cite{james01}.

Next, the trigger photon of the entangled pair is projected into an arbitrary polarization state with an adjustable strength polarizer~\cite{footnotepartialpolarizer} to remotely prepare a qubit of the form (\ref{arbitqubit}).  For perfect wave plates (i.e., the birefringent retardance is $180^\circ$ for a HWP and $90^\circ$ for a QWP), the precise wave plate angles can be readily calculated, similar to the case of \emph{directly} preparing arbitrary states~\cite{note:settings}.  If the wave plate phases deviate much from the ideal (e.g., when using nominally 702-nm wave plates for 670-nm photons), the precise wave plate orientations can be found numerically.  In this case, we maximize the fidelity between the state we wish to remotely prepare and the expected remotely prepared state calculated using the experimentally measured initial two-qubit entangled density matrix and the measured wave plate retardances.
\begin{figure}
\begin{center}
\includegraphics[width=8.6cm]{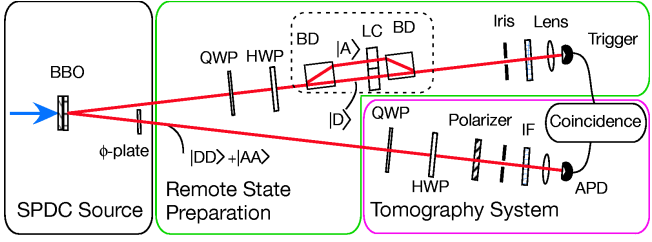}
\vspace{-.5cm}
\caption{Experimental arrangement for remote state preparation.   The entangled state $(|DD\rangle+|AA\rangle)/\sqrt{2}$ is generated by equally pumping two BBO crystals whose optic axes are in perpendicular planes [the relative phase is adjusted by tipping a HWP ($\phi$-plate) about its vertical optic axis].  The trigger photon is then {\em partially} projected into an arbitrary polarization state with a quarter-wave plate (QWP) and a half-wave plate (HWP) located before a partial polarizer~\cite{footnotepartialpolarizer} shown in the dashed box.  Conditional on detection of this photon, the sister photons is prepared into the desired state.}
\vspace{-1.7cm}
\label{rspexp}
\end{center}
\end{figure}
\begin{figure}
\begin{center}
\includegraphics[width=8.6cm]{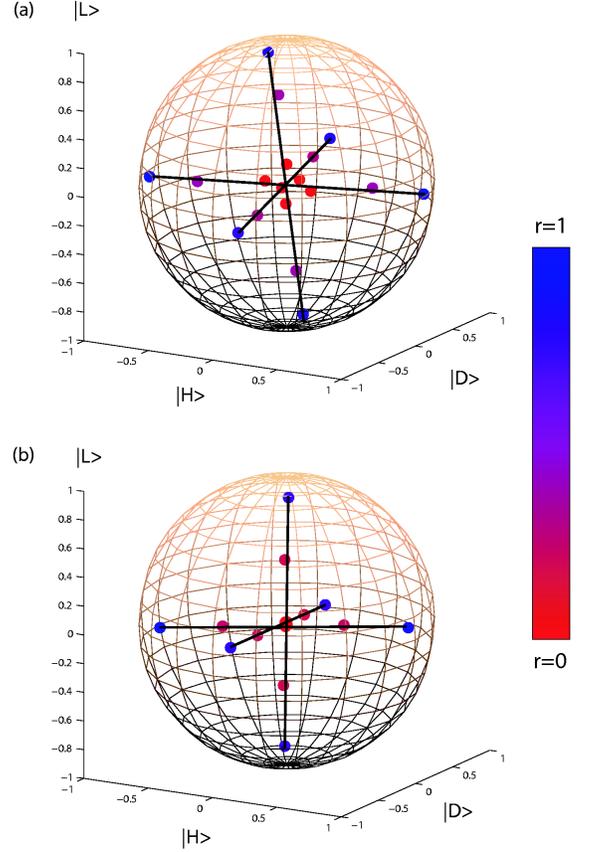}
\vspace{-0.75cm}
\caption{Remotely prepared states shown in the Poincar\'e sphere.  (a) States remotely prepared at 702 nm using frequency {\it degenerate} entanglement.  (b) States remotely prepared at 670 nm (using a 737-nm trigger).  In either case, the distance of the remotely prepared state from the origin is indicated by its color:  red $\rightarrow$ mixed, blue $\rightarrow$ pure.  Lines are drawn along the data to guide the eye.}
\vspace{-0.75cm}
\label{rspdat}
\end{center}
\end{figure}

The strength of the partial polarizer is determined by 
the transmission of two orthogonal polarization components, $T_D$ 
and $T_A$, each normalized with respect to each other by ${\it N}\equiv1/(T_D+T_A)$.  Unit 
transmission of one component, coupled with zero transmission of the 
other, is equivalent to an ideal polarizer for the transmitted component.  In contrast, if both transmitted components have the same amplitude, then the partial polarizer behaves as if no polarizer is present (though the overall amplitude may be reduced).  The action of the partial polarizer alone remotely prepares qubits according to~\cite{jpk04}~
\begin{equation}
\begin{array}{rcl}
\rho_{rp}{(T_D,T_A)} & = & N(T_D\langle D_t|\phi^+\rangle 
\langle\phi^+|D_t\rangle+ \\ & &  T_A\langle A_t|\phi^+\rangle \langle\phi^+|A
_t\rangle )\\
 & = & \frac{N}{2}(T_D|D_{rp}\rangle \langle D_{rp}|+T_A|A_{rp}\rangle
 \langle A_{rp}|) \\
& = & \frac{1}{2}\left( \begin{array}{cc}1 & N(T_D-T_A) \\
N(T_D-T_A) & 1 \end{array}  \right), 
\label{rhorp} 
\end{array}
\end{equation}
where the final density matrix is in the $|H\rangle,~|V\rangle$ basis.  After the partial polarizer, the trigger photons pass via a 2.2-mm iris, an interference filter (discussed below), and a collection lens, which focuses them onto a photon-counting avalanche photodiode (APD; Perkin-Elmer SPCM-AQR-13).  The classical communication from Alice to Bob is implemented by counting the RPQs in coincidence (within a 4.5-ns window) with their triggering photons.  The requirement for coincidence counting (which necessarily requires the (sub)luminal transfer of the APD signals) precludes all possibility of superluminal communication.

The final verification step is the tomographic measurement of the RPQ.  Using the wave plates and polarizer of the tomography system shown in Fig.~\ref{rspexp}, the remotely prepared ensemble is projected into  $\langle H|$, $\langle V|$, $\langle D|$ and $\langle A|$ states, as well as the left and right circular polarization states, $\langle L|\equiv\langle H|-i\langle V|$ and $\langle R|\equiv\langle H|+i\langle V|$, respectively.  The results of this complete polarization analysis are converted to the closest physically valid density matrix using a maximum likelihood technique~\cite{james01}.  

A summary of states remotely prepared in this way is shown in Fig.~\ref{rspdat}(a), along with a color bar indicating the distance of each RPQ from the center of the Poincar\'e sphere; the color corresponds to the state purity, from blue (pure) to red (mixed).  We tested our ability to precisely remotely prepare arbitrary states by creating six states along each of three (nearly orthogonal) axes in the Poincar\'e sphere.  To calculate experimental agreement between the state we prepared ($\rho_p$) and the remote state we expected ($\rho_e$) given the parameters of our system, we use the fidelity $F(\rho_e,\rho_p)\equiv \left| {\rm Tr}\left(\sqrt{\sqrt{\rho_e}\rho_p\sqrt{\rho_e}}\right) \right|^2$~\cite{jozsa94}; $F=$1~(0) for identical (orthogonal) states.  The average fidelity for our data is 0.996, with all 18 states above 0.99.  

The previous results were all taken using {\it degenerate} qubits, i.e., both trigger and RPQ were at 702 nm, as defined by the cut of the BBO crystals, the position of the collection irises (corresponding to a 3$^{\circ}$-opening angle with respect to the pump beam), and 2-nm bandpass filters in front of each APD~\cite{footnote:filter}.  To demonstrate the ability to remotely prepare qubits at other wavelengths, we additionally performed a similar set of measurements using {\it non}-degenerate entangled pairs:  Detection of a trigger photon after a 5-nm bandpass filter at 737 nm corresponded to a RPQ at 670 nm.  Note that all of the same physical resources, e.g., the crystals, the wave plates (by calculating wave plate phases away from design wavelengths) and the partial polarizer, were used at the different wavelengths.  Results are shown in Fig.~\ref{rspdat}(b).  The average fidelity was 0.996, with 17 of the 18 measured states above 0.99.  The flexibility to remotely prepare qubits at various wavelengths could be useful, e.g., for optimizing detector sensitivity, fiber or atmospheric transmission, or coupling to other quantum systems.  One could even envision a sort of nonlocal wavelength division multiplexing scheme: using an adjustable filter before the trigger detector, arbitrary states could be remotely prepared at one of several detectors, each receiving a slightly different wavelength band.

While a maximally entangled state resource enables the remote preparation of {\it any} state, it is important to consider the limits on the remotely preparable states when the two-qubit resource is mixed or only partially entangled, as in practice all realizable states are of this type.  We consider the scenario that Bob simply keeps or discards his photon, based on transmission of a single classical bit from Alice. Furthermore, we restrict Alice to single-qubit operations, i.e., no collective
manipulation of her qubits. This consideration is realistic, as efficient optical CNOT gates do not yet exist.

The most general operations Alice can perform on her qubit can be described by at most four
local filters~\cite{nc00}:
\begin{equation}
\rho_A\rightarrow \sum_{i=1}^4 p_i \,{\cal M}_i\rho_A{\cal M}_i^\dagger,
\end{equation}
where $\sum_i p_i \,{\cal M}_i^\dagger{\cal M}_i\le \openone$, and
each local filter ${\cal M}_i$ can be expressed in the singular-value 
decomposition
\begin{equation}
\label{eqn:dVDU}
{\cal M}=V^\dagger {\cal D} U.
\end{equation}
Here ${\cal D}$ is a non-negative, no-greater-than-unity diagonal matrix, and $U$ and $V$ are unitary matrices, not necessarily adjoint to each other.

Under the general operation (which can be non-trace preserving) performed by Alice, the initial joint two-qubit
state $\rho_{AB}$ becomes 
\begin{equation}
\label{eqn:POVM}
\rho_{AB}'=\sum_{i=1}^4 p_i \,{\cal M}_i\otimes\openone
\,\rho_{AB}\,{\cal M}_i^\dagger\otimes\openone,
\end{equation}
neglecting normalization, and Bob's qubit becomes $\rho'_B={\rm Tr}_A \rho_{AB}'$.
Thus, the most general states Alice 
can remotely prepare are mixtures of states which she can prepare
from a single local filter.  The states preparable from a single filter form a surface inside the Poincar\'e sphere, and all the states
she can remotely prepare lie on or inside the convex hull of this 
surface.  We now analyze the capability of
a general local filter applied to RSP.

The decomposition of a general local filter~(\ref{eqn:dVDU}) can be interpreted
as a three-step procedure: (i)~first, apply a local unitary transformation $U$, followed by (ii)~a ``Procrustean'' operation~\cite{bennett96a,bennett96b,kwiat01,peters04}
\begin{equation}
{\cal D}=\begin{pmatrix}
a & 0 \cr
0 & b
\end{pmatrix},
\label{eqn:d}
\end{equation}
with $0\le (a, b) \le 1$,
and lastly by (iii)~another unitary transformation $V^\dagger$.
The last step $V^\dagger$ has no effect on Bob's state, so it can
be ignored in the analysis of RSP. 
With a suitable parameterization of $U$, e.g., $U=\cos\theta\openone+i\hat{n}\cdot\vec{\sigma}\sin\theta$, where $\hat{n}$ is a unit vector and $\vec{\sigma}$ are the Pauli spin matrices,
it is straightforward to
analyze the states that can be remotely prepared by a single filter:
\begin{equation}
\label{eqn:Bob}
\rho_B={\rm Tr}_A\Big[ ({\cal D} U)\otimes \openone\, \rho_{AB} \,({\cal D} U)^\dagger\otimes \openone\Big],
\end{equation}
where $\rho_{AB}$ is the initial shared two-qubit state (unnormalized).

To illustrate the results, we first consider the case where $\rho_{AB}$ is a pure (but non-maximally) entangled state: $\sqrt{p}\ket{00}+\sqrt{1-p}\ket{11}$, assuming $p>1/2$ without loss of generality.  In fact, analysis of Eq.~(\ref{eqn:Bob}) reveals that with this state Alice can prepare {\it arbitrary} single-qubit states for Bob.  She first uses Procrustean distillation ${\cal D}$~\cite{bennett96a,bennett96b,kwiat01,peters04}, with  $a=\sqrt{(1-p)/p}$ and $b=1$,  to
obtain the perfect Bell state $\ket{00}+\ket{11}$ (though with probability $<1$), with which she may remotely prepare arbitrary states as we have demonstrated experimentally. 

As a rather different example, consider the starting state to be of the form~\cite{hhh97}
\begin{equation}
\rho_{AB}=\frac{1}{4}\left(\openone\otimes\openone+t_1\,\sigma_x\otimes\sigma_x+t_2\,\sigma_y\otimes\sigma_y+t_3\,\sigma_z\otimes\sigma_z \right),
\label{tet}
\end{equation}
which has eigenvalues $\lambda_1=(1 - t_1 +t_2 + t_3)/4$, $\lambda_2=(1 + t_1 - t_2 +t_3)/4$, $\lambda_3=(1 + t_1 + t_2 - t_3)/4$, and $\lambda_4=(1 - t_1 - t_2 - t_3)/4$.  This state, when described by $(t_1,t_2,t_3)$, lies on the surface of or inside
a tetrahedron, with the four vertices being
$(-1,-1,-1)$, $(-1,1,1)$, $(1,-1,1)$, and $(1,1,-1)$.  The state is entangled if any of the $\lambda_i$ are greater than 1/2.  Eq. (\ref{tet}) describes a wide range of interesting resource states, as judicious choice of $t_i$ changes the state from a maximally entangled pure state to a Werner state~\cite{werner} to states with varying classical correlations.  Again analyzing Eq.~(\ref{eqn:Bob}) we obtain that
the states (in the Poincar\'e sphere) that Alice can remotely prepare lie
 on or inside the ellipsoid centered at the origin, with three axes of length $|t_1|$, $|t_2|$, and $|t_3|$. To remotely prepare states on
 the surface of the ellipsoid, Alice simply rotates her qubit via
 $i \hat{n}\cdot\vec{\sigma}$, followed by projection onto $\ket{0}$. As she
 varies the rotation axis $\hat{n}=(\sin\alpha\cos\beta,\sin\alpha\sin\beta,\cos\alpha)$, Bob's states will then follow the corresponding trace $(t_1\sin2\alpha\cos\beta,t_2\sin2\alpha\sin\beta,t_3\cos2\alpha)$ on the ellipsoid.  To obtain states inside the ellipsoid, the projection onto $|0\rangle$ is replaced by the more general partial projection (Eq.~(\ref{eqn:d})).
  
We have seen that pure-state entanglement allows remote preparation of 
arbitrary states.  However, pure-state entanglement may not by {\it required} to remotely prepare some states.  Consider that the tetrahedron state (\ref{tet}) has purity 
\begin{equation}
P_{AB}={\rm Tr}\rho_{AB}^2=\frac{1}{4}(1+t_1^2+t_2^2+t_3^2),
\end{equation}
and is unentangled if $(t_1,t_2,t_3)$ lies inside the octahedron
embedded in the tetrahedron~\cite{hhh97}.  The maximum purity of the states Alice can remotely prepare via the tetrahedron state is 
\begin{equation}
\max P_{B}=\frac{1}{2}(1+ \max(t_1^2, t_2^2, t_3^2)).
\end{equation}
Interestingly, there appears to be no general requirement for entanglement in the two-qubit resource to be able to remotely prepare a one-qubit state of arbitrary purity.  Consider the classically correlated state $\rho_{cc}=\frac{1}{2}(\ketbra{00}+\ketbra{11})$ (i.e., $t_1=t_2=0$ and $t_3=1$).  This classically correlated two-qubit state can be used to remotely prepare any state of the form $\cos^2\theta\ketbra{0}+\sin^2\theta\ketbra{1}$, possessing any purity.  For unentangled resources where the classical correlations are less than in $\rho_{cc}$, Alice can only remotely prepare states near the origin of the Poincar\'e sphere.

We have demonstrated the first arbitrary remote state preparation of qubits, preparing a broad range of states spanning the Poincar\'e sphere.  The experimental methods employed may facilitate state control in linear optics feedforward quantum computation~\cite{rb01,walther05}.  Moreover, by varying the acceptance wavelength of the trigger photon (using a nondegenerate entangled source) we can also control the wavelength of the remotely prepared qubit.  Such a capability may assist in the preparation of states at wavelengths more optimal for other quantum communication protocols, e.g., quantum cryptography.  Finally, we have derived bounds on the single-qubit states that may be remotely prepared using certain two-qubit resource states.  

This work was supported by the U.~S.~Army Research Office (Grant No. DAAD19-03-1-0282).
\vspace{-0.40cm}
\bibliography{RSP}
\vspace{-0.40cm}
\end{document}